\documentclass[preprint,12pt]{elsarticle}

\usepackage{graphicx}
\usepackage{hyperref}

\journal{Comput. Phys. Commun.}

\begin{document}


\begin{frontmatter}

\title{{\tt JaxoDraw}: A graphical user interface for drawing Feynman diagrams.
Version {\tt 2.0} release notes.}

\author{D. Binosi}
\ead{d.binosi@gmail.com}
\address{ECT*, Villa Tambosi, Strada delle Tabarelle 286,
    I-38050 Villazzano (Trento), Italy}

\author{J. Collins}
\ead{collins@phys.psu.edu}
\address{Physics Department, Pennsylvania State University, USA}

\author{C. Kaufhold}
\ead{jaxodraw@chka.de}
\address{Striekenkamp 89, 28777 Bremen, Germany}

\author{L.Theussl}
\ead{ltheussl@gmail.com}
\address{Niels Bohr Institute, Copenhagen University,
    Blegdamsvej 17, Copenhagen 2100, Denmark}

\begin{abstract}
A new version of the Feynman graph plotting tool {\tt JaxoDraw} is presented.
Version {\tt 2.0} is a fundamental re-write of most of the {\tt JaxoDraw} core and some
functionalities, in particular importing graphs, are not backward-compatible
with the {\tt 1.x} branch. The most prominent new features include: drawing of
B\'ezier curves for all particle modes, on-the-fly update of edited objects,
multiple undo/redo functionality, the addition of a plugin infrastructure,
and a general improved memory performance. A new LaTeX style file is presented
that has been written specifically on top of the original {\tt axodraw.sty}
to meet the needs of this this new version.
\end{abstract}


\end{frontmatter}


\pagebreak

{\bf PROGRAM SUMMARY}\\
{\it Title of program}: {\tt JaxoDraw}\\
{\it Distribution format}: gzipped tar archive\\
{\it Operating system}:\\ 
Any Java-enabled platform, tested on Linux, Windows XP, Mac OS X
{\it Programming language used}: Java\\
{\it License}: GPL\\
{\it Catalogue identifier of previous version}: ADUA\\
{\it Journal Reference of previous version}: Comput. Phys. Commun. 161 (2004) 76--86\\
{\it Does the new version supersede the previous version?}: Yes\\
{\it Nature of problem}:\\ Existing methods for drawing Feynman diagrams usually
require some `hard-coding' in one or the other programming-
or scripting language. It is not very convenient and often time consuming, to
generate relatively simple diagrams.\\
{\it Method of solution}:\\ A program is provided that allows for the interactive
drawing of Feynman diagrams with a graphical user interface. The program is easy
to learn and use, produces high quality output in several formats and runs
on any operating system where a Java Runtime Environment is available.\\
{\it Reasons for the new version}:\\ A variety of new features and bug fixes. \\
{\it Summary of revisions}:\\ Major revisions since the last published user
guide were versions {\tt 1.1}, {\tt 1.2} and {\tt 1.3} with several minor
bug-fix releases in between.\\
{\it Restrictions}:\\ To make use of the latex export/preview functionality,
a latex style file has to be installed separately.
Certain operations (like internal latex compilation,
Postscript preview) require the execution of external commands that might not
work on untested operating systems.\\
{\it Typical running time}: As an interactive program, the running time depends
on the complexity of the diagram to be drawn.

\bigskip


\pagebreak

{\bf LONG WRITE-UP}

\section{Introduction}
\label{intro}

Since the first public release of {\tt JaxoDraw-1.0}~\cite{Binosi:2003yf}
in September 2003, the program has been continuously improved by the authors,
partly motivated by the need of correcting obvious flaws and adding relevant
new features, but above all by the overwhelming user
feedback and encouragement by the community. During the last five years,
{\tt JaxoDraw} has become one of the most popular tools to draw Feynman
diagrams, as witnessed by the number of citations and acknowledgements
in scientific papers, as well as positive recommendations in public
physics forums and reviews. The program is now included and packaged
by default by various Linux distributions and has been incorporated
into large-scale physics applications like the
{\tt jHepWork}~\cite{Chekanov:2008za} framework.

Version {\tt 2.0} is a major upgrade which
has seen a fundamental re-write of most of the {\tt JaxoDraw} core with
respect to the {\tt 1.x} line of development. The latter had already produced
some major upgrades (versions {\tt 1.1}, {\tt 1.2} and {\tt 1.3})
with several minor bug-fix releases in between.

The current document only outlines the major changes with respect to the
features described in the published User Guide~\cite{Binosi:2003yf} for
{\tt JaxoDraw-1.1}. The complete User Guide for the current version is
included in the program and may be consulted on the {\tt JaxoDraw} web site,
see sec.~\ref{info} for some links.

Please refer to~\cite{Binosi:2003yf} and to the present paper if using version {\tt 2.0} of the program. 

\subsection{Overview of main new features}

The following list gives a quick overview of the most prominent
new features:

\begin{itemize}

\item Added a plugin infrastructure. Users can write plugins for custom
        import/export formats that can be installed independently and will
        be recognized by new versions of {\tt JaxoDraw}.

\item Added B\'ezier curves as drawing style. B\'eziers can be drawn for all
        particle modes, including gluons and photons.

\item The dimensions and positions of arrows can be adjusted.

\item Added scroll bars to the drawing area. Now if an object is resized
        or moved beyond the current canvas size, scroll bars will appear.

\item Added multiple undo/redo functionality. The user is not bound to a
        single undo operation, but can undo (and redo) a number of steps.

\item Editing objects from the edit panel now has an immediate effect on the
        object so that editing operations can be previewed on the fly.

\end{itemize}

\begin{figure}[ht]
\includegraphics[width=\textwidth]{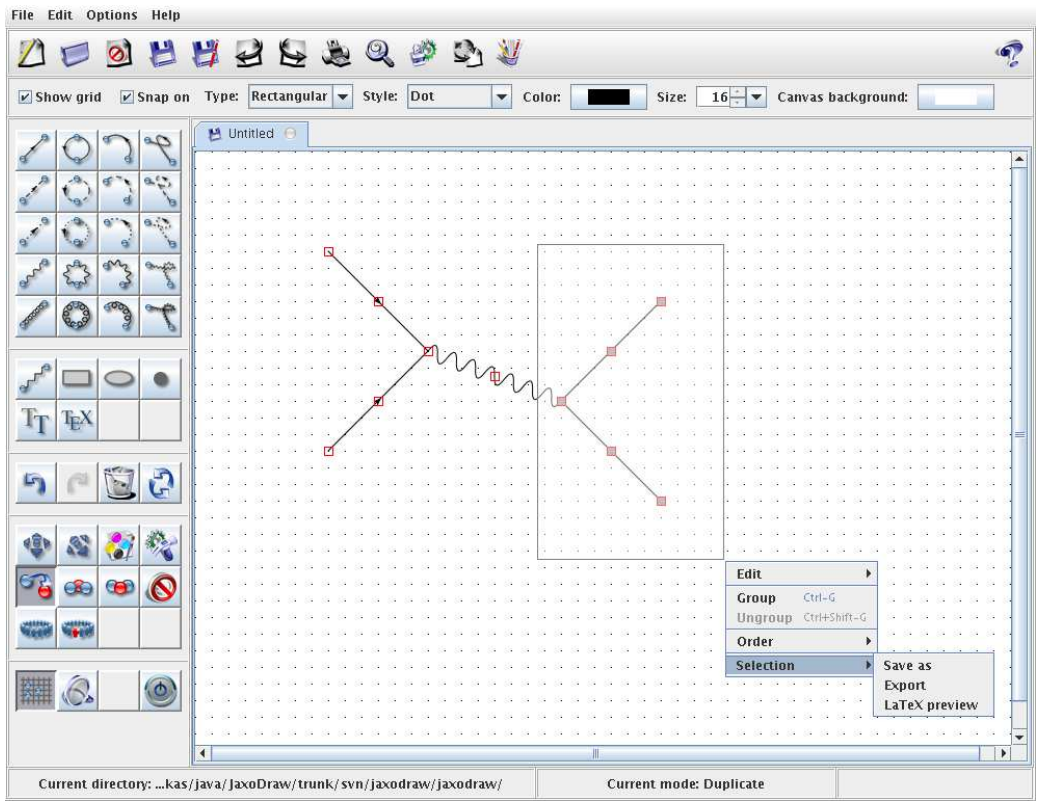}
\caption{{\tt JaxoDraw} in action.}
\label{jaxoscreen}
\end{figure}

\noindent
{\em Due to extensive refactoring, {\tt JaxoDraw-2.0} is not compatible
with any earlier version of the program. In particular,
xml files that were generated with earlier versions will generally
not be imported correctly.}


\section{Changes between v. 1.3-2 and v. 2.0}
\label{changes20}

In this section we list the main changes that happened between the last released
version of {\tt JaxoDraw} and the new version {\tt 2.0}.

\paragraph{General}

\begin{list}{$\bullet$}{}
\item {\tt JaxoDraw} now uses its own version of the axodraw style file,
        called {\tt axodraw4j.sty}. This style file has to be installed separately.
        See App.~\ref{axodraw4j} for a description of {\tt axodraw4j.sty}.
\item Version {\tt 2.0} has seen a number of fundamental changes in the underlying
        class structure of {\tt JaxoDraw}. These changes were motivate mainly
        by performance and maintainability arguments, but it implies a certain
        backward incompatibility as graphs that where saved/exported with older versions
        of {\tt JaxoDraw} will not be opened/imported correctly anymore.
\item Improved memory performance which facilitates working with several graphs
        at the same time.
\item Several key strokes have been set to more ``standard" combinations.
        These include: Copy ({\tt Ctrl+C}), Paste ({\tt Ctrl+V}), Cut ({\tt Ctrl+X}),
        Save ({\tt Ctrl+S}), Save As ({\tt Ctrl+Shift+S}), Import ({\tt Ctrl+R}),
        Export ({\tt Ctrl+ Shift+R}).
\item The ``Copy" mode has been renamed to ``Duplicate" to avoid confusion
        with the usual clipboard ``Copy" action.
\item SVG support has been removed from the core program, it is now
        available as a separate plugin.
\item The preferences are not stored in a file {\tt .Jaxorc} anymore,
        the standard Java Preferences API is used instead.
        See Sec.~(\ref{config-dir}) for details.
\item The User Guide is not bundled in the jar archive anymore,
        but created at run-time and installed into a local directory.
        See Sec.~(\ref{user-guide}) for details.
\item Log records are now written to a rotating sequence of files
        in a default log directory. See Sec.~(\ref{logging}) for details.
\item The drawing of photons and gluons has been adjusted to use the same
        algorithms as the postscript code in the {\tt axodraw4j} style file.
        Different output formats are entirely consistent now, however
        some strange effects happen for special configurations,
        e.g. gluon loops with very small radii.
\end{list}

\paragraph{New features}

\begin{list}{$\bullet$}{}
\item Added a plugin infrastructure. Users can write plugins for custom
        import/export formats that can be installed independently and will
        be recognized by new versions of {\tt JaxoDraw}.
        See Secs.~(\ref{using-plugins},\ref{writing-plugins}) for details.
\item Added B\'ezier curves as drawing style. B\'eziers can be drawn for all
        particle modes, including gluons and photons.
\item Photon and gluon arcs are now painted during resizing.
\item Make the symmetry of a photon arc configurable.
\item Editing objects from the edit panel now has an immediate effect on the
        object so that editing operations can be previewed on the fly.
\item Added the new feature that a double click with the right mouse button on
        the canvas brings up the edit panel for the nearest object.
\item The faint box method for selecting objects via a right click and drag action on the Canvas has been improved. 
         On button release, a menu will be presented with a choice of
        operations applicable to the current selection of enclosed objects. 
        In particular the selected objects can be previewed, exported and/or saved.
\item Added scroll bars to the canvas. Now if an object is resized
        or moved beyond the current canvas size, scroll bars will appear.
\item Added multiple undo/redo functionality. The user is not bound
        to a single undo operation, but can undo (and redo) a
        number of steps determined by the {\tt Undo depth} preference 
        (currently limited to $10^4$ moves).
\item Groups can be rescaled by dragging, like all other objects.
\item Objects with fill color can now be optionally unfilled (only drawing their boundary).
\item A grid bar has been added to the main panel which shows at a glance all the grid settings of the current tab.       
\item The grid can now be customized. In particular several grid styles are now available: {\tt dot}, {\tt cross}, {\tt line} 
          and {\tt honeycomb} (the latter being available only when the grid type is set to {\tt hexagonal}).
\item Holding down the mouse button after a center click on the canvas remove
          all handles that might be cluttering a graph in Edit mode. 
\item Arrows can be customized and arbitrarily positioned. The parameters that can be controlled 
          are the arrow {\tt length}, {\tt width} and {\tt inset}. 
          This works also with \LaTeX/\LaTeX{\tt ->EPS} export, thanks to the new sty file {\tt axodraw4j}.
\item Different exporting formats have now different options.
\item Warning system when exporting with unsupported options.
\item Color space can be set by the user. The available options are {\tt Axodraw} 
         (coinciding with the usual {\tt colordvi} color space) and {\tt complete} (representing the complete RGB space).
         When working with the complete color space and exporting to \LaTeX/\LaTeX{\tt ->EPS}, color conversion 
         will be applied.
\item Edited objects are automatically brought in the forefront, for better visualization of the editing operations. 
       After the finalization of the editing operations, they will go back to the original ordering.
\item Edit panels are brought up in positions that avoid covering the edited object as much as possible. 
\item A lot of new preferences can be now set via the Preferences panel.
\item The {\tt Copy} action from different Tabs has been streamlined, through the use of the faint box method.
\item Multiple files can be specified on the command line
         and will be opened in multiple tabs.
\item The command line option {\tt -nosplash} can now be used in order not to
         show the splash window on startup.
\item The command line option {\tt --convert} can now be used to convert a
      number of {\tt JaxoDraw} xml files to {\tt axodraw4j} tex files
      (and vice versa) without the need of bringing up the user interface.
\end{list}

\paragraph{Bug fixes}

\begin{list}{$\bullet$}{}
\item Fixed the bug where file operations (Open, Save, Import, ...) did not work
        on Windows if the path to the file contained whitespaces.
\item Fixed the bug that gluon loops did not close in latex output. This was a
        bug in {\tt axodraw.sty} that is fixed in the new {\tt axodraw4j.sty}.
\item Fixed the bug that double line separation and line width was defined
        inconsistently which led to different results in latex and postscript
        output.
\item Prevent various user input dialogs from going into the
        background and blocking the main window.
\item Fixed the bug that made {\tt JaxoDraw} hang if
        {\tt axodraw.sty} was not installed or not found.
\item The hexagonal grid strategy has been changed to make the grid uniform,
        not subject to rounding, which could make points drop from the grid
        if objects were moved.
\item A large number of minor bug fixes that are detailed in the {\tt CHANGES}
        document of the source distribution.
\end{list}


\section{New features since v. 1.1}
\label{newfeatures}

This section only lists the new features that were added at every major release
since version {\tt 1.1}, which is the version that was described in our first
published description of the program~\cite{Binosi:2003yf}.
Each of the releases below also incorporated a number of bug
fixes that can be tracked from the corresponding release notes. In addition,
a few point releases were also made ({\tt 1.0-1}, {\tt 1.3-1}, {\tt 1.3-2})
with only minor bug fixes.

\subsection{New features in v. 1.3}
\label{features13}

\begin{list}{$\bullet$}{}
\item Make the Mac OS X README file available from the Help menu (Mac only).
\item New vertex type diamond.
\item Implemented LaTeX text rotation (using the pstricks package) and rotation
        of Postscript texts.
\item Added default return mode.
\item New export/preview formats SVG, JPG and PNG.
\item Added a dynamic zoom.
\item Rewrote the export dialog for key-friendlyness: tab key toggles between
        items, space selects, escape cancels. In the export formats combobox
        you can choose an entry by pressing its first character or go up and
        down with the arrow keys.
\end{list}


\subsection{New features in v. 1.2}
\label{features12}

\begin{list}{$\bullet$}{}
\item Added a hexagonal grid. Each tab can have its own grid type and size.
\item Introduced a `WatchFile' mode to avoid opening new windows for each
        preview.
\item Introduced radio buttons in the vertex menus to indicate the currently
        active Vertex mode.
\item PSText now can display curly brackets like in LaTeX: $\backslash\{$
        and $\backslash\}$.
\item Several Mac OS X specific enhancements, eg. menu key short cuts,
        key short cuts for middle and right mouse button, a preference for
        the latex and dvips path, which allows internal latex compilation, etc.
\item Arcs and triangular vertices are now three-point objects: they are drawn
        with a click for each point.
\item The Preferences panel has been restructured for a clearer layout.
\item Many more fonts are now available in PSText mode because we do not filter
        out fonts anymore that cannot display greek characters.
\end{list}


\section{Guide to new features}
\label{guide}

This section describes in more detail some of the most prominent new features
of {\tt JaxoDraw-2.0} with respect to the last released version.

\subsection{Local configuration directory and preferences}
\label{config-dir}

The preferences are not stored in a file {\tt .Jaxorc} (in the user's home
directory {\tt \$USER\_HOME/}) anymore, the standard Java Preferences API
is used instead. A folder {\tt \$USER\_HOME/.jaxodraw/\$VERSION/}
(in the following called {\tt \$JAXO\_DIR}) is used to store
all program-specific information, currently there are sub-directories for
log files, plugins and the User Guide.

\subsection{User guide}
\label{user-guide}

The User Guide is not bundled in the distributed program (jar archive) anymore,
but created at run-time and installed into {\tt \$JAXO\_DIR/usrGuide/}.
It can therefore be opened by any custom browser and be viewed locally and
independently of {\tt JaxoDraw}.

\subsection{Logging}
\label{logging}

Log records are now written to a rotating sequence of files in the default log
directory {\tt \$JAXO\_DIR/log/}. The logging level for the written log records
is always kept at {\tt DEBUG}, only the logging level for the console output can
be configured (e.g. with the {\tt --debug} or {\tt --quiet} command-line options).

\subsection{Plugins}
\label{using-plugins}

In version {\tt 2.0} a plugin architecture was added to {\tt JaxoDraw}.
Plugins are software components that may optionally be added to the program
at runtime, ie without the need of changing or re-compiling the main program.
This makes it easy to use optional features, in particuar export to uncommon
formats or import of other custom file formats, while keeping the size of the
main program at a minimum. In fact, the original main purpose of the plugin
architecture was to draw some functionality out of the {\tt JaxoDraw} core.

Plugins are installed (and un-installed) using the Plugin Manager panel which is
accessible from the Options menu. Once a plugin is installed, {\tt JaxoDraw}
will automatically recognize it at start-up and the corresponding functionality
will be available for the current session and subsequent sessions until the
plugin is uninstalled. Plugins are installed in {\tt \$JAXO\_DIR/plugins/}.

A list of available plugins is maintained on the {\tt JaxoDraw} web site,
currently there are plugins available for export to PDF (Portable Document Format)
and SVG (Scalable Vector Graphics) format.

\subsection{Writing custom import/export plugins}
\label{writing-plugins}

The most interesting consequence of the plugin architecture is that it allows
anybody to write custom plugins that may be loaded by {\tt JaxoDraw} and
used by anybody without re-compiling the main program.

As an illustration, imagine we have a Feynman diagram coded in some custom
input file (eg created by another program), and we would like to import this
diagram into {\tt JaxoDraw} so that it can be edited interactively.
All we have to do is write a Java class that extends {\tt JaxoImportPlugin}
and implement all the required abstract methods, the most important in
this case being {\tt importGraph}:

{\small
\begin{verbatim}
public class MyImportJaxoPlugin extends JaxoImportPlugin
{
    protected JaxoGraph importGraph(InputStream inputStream)
        throws JaxoPluginExecutionException
    {
        // return a graph read from an InputStream
    }
}
\end{verbatim}
}

A ready-compiled plugin can then be installed by the {\tt JaxoDraw} Plugin
Manager without changing anything in the main program. In particular, we can
publish the plugin so other people can install and use it as well, without
having to download and install a new version of {\tt JaxoDraw}!

For more detailed instructions and further information on
writing plugins please refer to the {\tt JaxoDraw} web site.


\section{Additional information}
\label{info}

The {\tt JaxoDraw} home page is at
\url{http://jaxodraw.sourceforge.net/}.
It contains up-to-date information about the program, in particular links to
our mailing
lists\footnote{\url{http://jaxodraw.sourceforge.net/mail-lists.html}}
and bug-tracking
system\footnote{\url{http://jaxodraw.sourceforge.net/issue-tracking.html}}.


\bigskip

{\bf Acknowledgements}

\noindent
We would like to thank all the people on the {\tt JaxoDraw} mailing
list for their help and feedback during the testing phase.


\begin{appendix}

\section{New features of {\tt axodraw4j}}
\label{axodraw4j}

{\tt JaxoDraw-2.0} uses a new \LaTeX~style file called {\tt axodraw4j.sty}
(i.e. '{\tt axodraw} for {\tt JaxoDraw}') for its LaTeX exports which is based
on J.~Vermaseren's original {\tt axodraw.sty}~\cite{Vermaseren:1994je}.
The name has been changed to avoid any backward compatibility issues with old
documents that use the original {\tt axodraw}.

We describe here, from a user's perspective, how {\tt axodraw4j.sty} differs
from the original version. The reader unfamiliar with {\tt axodraw} should
therefore first refer to the documentation for the original
version\footnote{\url{http://www.ctan.org/get/graphics/axodraw/axodraw.pdf}}.

\subsection{Main changes from axodraw to axodraw4j}

\begin{itemize}

\item Lines (solid, dashed, photon, and gluon) can now be made double,
    with an adjustable separation.

\item The dimensions and positions of arrows can be adjusted.

\item Lines and dashed lines can be made from B\'ezier curves.

\item Since there are now many more possibilities to specify a line,
    optional arguments to the main line drawing commands can be used
    to specify them in a keyword style.

\item A new macro named \texttt{$\backslash$Arc} is introduced for arcs
    and dashed arcs.

\item For consistency the \texttt{$\backslash$GlueArc} macro is renamed to
    \texttt{$\backslash$GluonArc}, with the old macro retained as a synonym.

\item Some bugs are corrected. The most notable one is that {\tt axodraw4j}
    now works correctly with {\tt revtex} and {\tt revtex4}.

\item The behavior of arcs is changed when the specified opening
    angle is outside the natural range.

\item The macros originally specified as \texttt{$\backslash$B2Text},
    \texttt{$\backslash$G2Text}, and \texttt{$\backslash$C2Text}
    are now named \texttt{$\backslash$BTwoText}, \texttt{$\backslash$GTwoText},
    and \texttt{$\backslash$CTwoText}.

\end{itemize}

\subsection{Commands}

The {\tt axodraw4j.sty} file has been kept largely backward compatible
with the original {\tt axodraw}, i.e. almost all commands that were
available in {\tt axodraw} are also implemented in {\tt axodraw4j}
(the exception being the renaming of commands like
\texttt{$\backslash$BTwoText}, etc.).  The few additional commands and the
enhanced old commands are documented below.  For a description of the
remaining commands, please refer to the original documentation of {\tt
  axodraw}~\cite{Vermaseren:1994je}.

A typical use of the macros in a document is as follows:

\begin{verbatim}
\documentclass{article}
\usepackage{axodraw4j}
\begin{document}
 \begin{picture}(162,39) (0,0)
   \Line[arrow,arrowlength=5,arrowwidth=2](0,19)(48,19)
   \Arc[arrow,arrowlength=5,arrowwidth=2](80,43)(40,143,36)
   \Line(112,19)(160,19)
   \GluonArc(80,-5)(40,37,143){3.5}{6}
 \end{picture}%
\end{document}
\end{verbatim}

The main macros for line drawing are \texttt{$\backslash$Arc},
\texttt{$\backslash$Bezier}, \texttt{$\backslash$Gluon},
\texttt{$\backslash$GluonArc}, \texttt{$\backslash$Line},
\texttt{$\backslash$Photon}, and \texttt{$\backslash$PhotonArc}.
In each of the following descriptions of the macros, the part enclosed
in square brackets, ``\texttt{[{\it options}]}", is an optional
argument, with the options being specified by keywords, as explained later.

\begin{itemize}

\item \texttt{$\backslash$Arc[\mbox{{\it options}}]($x$,$y$)($r$,$\phi_1$,$\phi_2$)}

  Draws a circular arc.  The center of the arc is ($x$,$y$), the
  radius is $r$, and the starting and ending angles are $\phi_1$ and
  $\phi_2$ (in degrees).  By default, the arc is an anticlockwise single
  solid line without an arrow.  

    Supported option groups are: {\tt arrow}, {\tt clock}, {\tt dash}, {\tt double}.

\item \texttt{$\backslash$Bezier[{\it options}]($x_1$,$y_1$)($x_2$,$y_2$)($x_3$,$y_3$)($x_4$,$y_4$)}

    Draws a B\'ezier line with control points
    ($x_1$,$y_1$), ($x_2$,$y_2$), ($x_3$,$y_3$), and ($x_4$,$y_4$).

    Supported option groups are: {\tt arrow}, {\tt dash}, {\tt double}.

\item \texttt{$\backslash$Gluon[{\it options}]($x_1$,$y_1$)($x_2$,$y_2$)$\{${\it
      amplitude}$\}\{${\it windings}$\}$}

  Draws a gluon from ($x_1$,$y_1$) to ($x_2$,$y_2$). The width of the
  gluon is twice the `amplitude' parameter, and the number of windings
  is the `windings' parameter, which is rounded down to an integer.
  The side at which the windings lie is determined by the order of the
  two coordinates. Also a negative amplitude changes this side.

    Supported option groups are: {\tt double}.

\item \texttt{$\backslash$GluonArc[{\it options}](x,y)(r,$\phi_1$,$\phi_2$)$\{${\it
      amplitude}$\}\{${\it windings}$\}$}

  Draws a gluon on a circular arc.  The center of the arc is
  ($x$,$y$), the radius is $r$, and the starting and ending angles are
  $\phi_1$ and $\phi_2$ (in degrees).  By default, the arc is anticlockwise.

  The width of the gluon is twice the `amplitude' parameter, and the
  number of windings is the `windings' parameter, which is rounded
  down to an integer.  Whether the curls are inside or outside depends
  on the sign of the amplitude. When it is positive the curls are on
  the inside.

    Supported option groups are: {\tt clock}, {\tt double}.

\item \texttt{$\backslash$Line[{\it options}]($x_1$,$y_1$)($x_2$,$y_2$)}

  Draws a line from ($x_1$,$y_1$) to ($x_2$,$y_2$).  By default the
  line is a solid single line without an arrow.

    Supported option groups are: {\tt arrow}, {\tt dash}, {\tt double}.

\item \texttt{$\backslash$Photon[{\it
      options}]($x_1$,$y_1$)($x_2$,$y_2$)$\{${\it amplitude}$\}\{${\it wiggles}$\}$}

  Draws a photon from ($x_1$,$y_1$) to ($x_2$,$y_2$).  The width of
  the gluon is twice the `amplitude' parameter, and the number of
  windings is the `windings' parameter.  The number of windings is
  rounded down to an integer or half integer. Whether the first wiggle
  starts up or down depends on the sign of the amplitude.  If the
  amplitude (rounded) is a half integer, the photon is symmetric.

    Supported option groups are: {\tt double}.

\item \texttt{$\backslash$PhotonArc[{\it options}](x,y)(r,$\phi_1$,$\phi_2$)$\{${\it
      amplitude}$\}\{${\it wiggles}$\}$}

  Draws a photon on a circular arc.  The center of the arc is
  ($x$,$y$), the radius is $r$, and the starting and ending angles are
  $\phi_1$ and $\phi_2$ (in degrees).  The width of the gluon is twice the
  `amplitude' parameter, and the number of windings is the `windings'
  parameter, which is rounded down to an integer.  By default, the arc
  is anticlockwise.

  The sign of the amplitude determines whether the photon starts going
  outside (positive) or starts going inside (negative). If the photon
  is to reach both endpoints from the outside the number of wiggles
  should be an integer plus 0.5.

    Supported option groups are: {\tt clock}, {\tt double}.

\end{itemize}

\subsection{Options}

Options are organized in logical groups for different kinds of
property, and are specified by keyword-value pairs, as in
\begin{verbatim}
   \Line[arrow=true,arrowlength=5,arrowwidth=2](0,19)(48,19)
\end{verbatim}
For a boolean keyword, the keyword alone is equivalent to the true
value. Before options are processed, from left to right, the values
are set to initial default values. For the boolean options, the
initial values are all false.

Each command only implements a subset of the keywords.  The possible
keywords, organized by groups, are
\begin{description}

\item[{\tt dash}] Boolean option specifying that the line is dashed.

\item[{\tt dashsize}] The size of the dashes. Defaults to $3$.

\item[{\tt double}] Boolean option specifying that the line is
  doubled.

\item[{\tt linesep}] The separation of a double line. It is the
  distance between the center of the two component single-lines.
  Defaults to 2.

\item[{\tt clock}] Boolean option specifying that an arc is drawn
  clockwise from the starting angle instead of anticlockwise. 

\item[{\tt arrow}] Boolean option specifying that an arrow is drawn on
  the line.

  If neither of the arrow dimensions (length or width) is specified,
  the (half-)width defaults to $1.2 * (2 + \texttt{linewidth})$ for a
  single line, and to $1.2 * (2 + 0.7 * \texttt{linesep} +
  \texttt{linewidth})$ for a double line, and the length defaults to
  $2.5 * \mbox{\tt arrow-half-width}$.  If only one of the width and length is
  specified, the other is determined by the same aspect ratio as for a
  default arrow.

\item[{\tt arrowpos}] The position of the arrow on an object (line /
  arc / loop / bezier).  It is given as a fraction between 0 and 1.
  Defaults to 0.5 (i.e., the arrow is at the middle of the object).

\item[{\tt arrowscale}] If neither an explicit arrow length nor an
  explicit arrow width is given for a line, then the value given with
  this keyword specifies the linear scale of the arrow relative to the
  default size of the arrow.

\item[{\tt arrowlength}] The length of the arrow.  

\item[{\tt arrowidth}] The half-width of the arrow.

\item[{\tt arrowinset}] The inset of the tail of an arrow.  It is
  specified as a fraction of the arrowlength, a number between 0 and
  1.  Defaults to 0.2.

\item[{\tt flip}] Boolean option specifying that the arrow is reversed
  (flipped) relative to the direction of the line, i.e., that it
  points from the end towards the start of the line.

\end{description}

Macros are available for setting certain parameters without the need
to specify them in individual line commands. In each of the
descriptions below, \texttt{num} represents a number

\begin{description}

\item[\texttt{$\backslash$SetArrowScale$\{${\it num}$\}$}]
   Sets the scale of an arrow relative to the default.  This is used
   on a line with an arrow when the dimensions are not otherwise
   specified. Its value is initialized to 1.

\item[\texttt{$\backslash$SetArrowInset$\{${\it num}$\}$}]
   Sets the inset of the tail of an arrow relative to its length.
   This is used on a line with an arrow when the arrowinset is not
   explicitly specified.  Its value is initialized to 0.2.

\item[\texttt{$\backslash$SetArrowAspect$\{${\it num}$\}$}]
   Sets the default
   aspect ratio of arrows; this is the ratio of the length of an arrow
   to its \emph{full} width.  This is used in arrows when no
   dimensions are specified or when only one of the length and width
   are specified.  Its value is initialized to 1.25.

\item[\texttt{$\backslash$SetArrowPosition$\{${\it num}$\}$}]
   Sets the default position of arrows along a line: 0 is
   at the start, 1 is at the end, 0.5 is at the center.  This is used
   on a line with an arrow when the arrowpos option is not used.  Its
   value is initialized to 0.5.

\end{description}

\end{appendix}




\end{document}